\documentstyle[12pt,aasms4]{article}
\newcommand{\keV}{ke\kern-.15emV}

\begin{document}

\title{Re-observing the EUV emission from Abell 2199: {\it in situ}
measurement of background distribution by offset pointing}

\author{Richard~Lieu$\,^{1}$, Massimiliano Bonamente$\,^{1}$,
Jonathan~P.~D.~Mittaz$\,^{2}$, Florence Durret$\,^{3}$, and
Sergio Dos Santos$\,^{3}$ and Jelle S. Kaastra$\,^{4}$}

\affil{\(^{\scriptstyle 1} \){Department of Physics, University of Alabama,
Huntsville, AL 35899, U.S.A.}\\
\(^{\scriptstyle 2} \){Mullard Space Science Laboratory, UCL,
Holmbury St. Mary, Dorking, Surrey, RH5 6NT, U.K.}\\
\(^{\scriptstyle 3} \){Institut d'Astrophysique de Paris, CNRS,
98bis Bd Arago, F-75014 Paris, France}\\
\(^{\scriptstyle 4} \){SRON Laboratory for Space Research,
Sorbonnelaan 2, NL-3584 CA Utrecht, The Netherlands}\\
}

\begin{abstract}
The EUV excess emission from the clusters A2199 and A1795 remains an
unexplained astrophysical phenomenon.  There has been many
unsuccessful attempts to `trivialize' the findings.  In this
Letter we present direct evidence to prove that the most
recent of such attempts, which attributes
the detected signals to a background non-uniformity effect, is
likewise excluded.  We address
the issue by a re-observation of A2199 which features a new filter
orientation, usage of a more sensitive part of the detector and, crucially,
includes a background pointing at $\sim$ 2$^o$ 
offset - the first {\it in situ}
measurement of its kind.  We demonstrate 
quantitatively two facts: (a) the offset pointing provides an accurate
background template for the cluster observation, while (b) data from
other blank fields do not.
We then performed point-to-point subtraction of the
{\it in situ} background from the cluster field,
with appropriate propagation of errors.  The
resulting cluster radial profile is consistent with that obtained by
our original method of
subtracting a flat asymptotic background.  The emission now extends to a radius
of 20 arcmin; it confirms the rising prominence of EUV excess beyond $\sim$ 5
arcmin as previously reported.
\end{abstract}

\noindent
{\it Subject headings:} Galaxies: clusters: general; instrumentation:
detectors; methods: data analysis, statistical; radiation mechanisms:
thermal.
  
The origin of diffuse EUV and soft X-ray emission from the Virgo and Coma
clusters of galaxies (Lieu et al 1996a,b), detected at a level higher than that
expected from the hot intracluster medium (i.e. the cluster soft excess, or
CSE, syndrome), has remained unsolved.  The most
serious theoretical puzzle is presented by the EUVE data of the
rich clusters Abell 1795 and 2199, where the CSE emission was found to be very
soft (with luminous EUV excess unaccompanied by any excess in the 1/4-keV band
of the ROSAT PSPC) and absent from the cluster centers (Mittaz, Lieu \& Lockman
1998; Lieu, Bonamente \& Mittaz 1999, hereafter abbreviated as LBM).
The reported phenomena have profound implications irrespective of whether
the emission turns out to be thermal, non-thermal, or some other origin (Cen \&
Ostriker 1999, Sarazin 1999, Lieu et al 1999 and references therein).
It is therefore not surprising that questions
concerning the observational integrity of CSE are still occasionally raised,
especially with respect to A1795 and A2199.  The area of data
analysis being scrutinized recently is the subtraction of the EUVE detector
background, as the inferred signals at large cluster radii, being only a small
fraction of the background, are sensitive to this procedure.

To understand the potential problems involved in analysing data from the EUVE
DS detector, we first describe the essential aspects of the DS background
behavior which can affect the analysis of cluster data.  We emphasize that here
and after, unless otherwise specifically stated, only raw data (i.e. the output
product of the EUVE standard telemetry processing pipeline, as publically
archived) have been employed.  The formidable problems which confront usage of
additionally manipulated data will be enlisted shortly.  

The portion of the DS occupied by the Lex/B (69 - 190 eV) filter is rectangular
in shape, and the sensitivity of the detector to background photons and
particles is not spatially uniform across this area\footnote{Note that this
sensitivity non-uniformity is {\it not} a vignetting effect, due to the obvious
lack of azimuthal symmetry around boresight}.  In Figure 1 we show contours of
the background as obtained by accumulating detector images of three
extragalactic pointings, viz. 2EUVE J1100+34.4, 2EUVE J0908+32.6, and M15,
totalling an exposure of $\sim$ 85 ksec, comparable to the longest exposure
EUVE has in any single observation of an individual cluster.  All bright
sources have been removed.  The figure results from a multi-scale wavelet
analysis and reconstruction (see Slezak, Durret \& Gerbal 1994) of the detector
image, retaining features (in every spatial scale) which have a minimum
significance level of 3 $\sigma$.  It can be seen from Figure 1 that there is
no evidence of extended emission at any scale, and that the background exhibits
large scale gradients at the extremeties of the y-axis.  Note that this spatial
pattern is typical of the DS Lex/B background for commensurate exposures.
Thus, although such a 2-D view offers firm assurance that cluster glows are
inherently not present in background fields, it also reveals the potential
dangers of a 1-D view, viz. that any radial surface brightness profile centered
at or near boresight (where clusters are usually observed) could lead to an
underestimate of the background if this is taken from an annulus with large
(inner and outer) radii which encompasses the bands of lower background.

Nonetheless, a simple way of minimizing this difficulty does exist, and
(though not explicitly stated) has been the manner in which analysis leading to
our past publications was performed.  The same procedure is also
adopted in the present work.
It involves truncating areas of large and
small y coordinate.  There are no strict criteria on how to execute this,
although our approach is to first plot the total
detector counts as a function of y after they have been summed over all values
of x, and then place the y-limits at the extreme ends of the `plateau' region.  
To illustrate, we show in
Figure 2 the y-profile of the composite detector image of many long 
exposure
fields (Mrk 421, NGC 5548, PKS 2155-304, PSR J0108-1431, and the 
forementioned
three fields, totalling an exposure of 0.736 Msec). 
It will be shown below, using the current data as example, that
the method leads to a well-behaved background profile.

Other potential issues are the use of a template background and the effect of
data manipulation.  Detailed treatment of this subject is given in Kaastra
(1999) but we summarize the main points here.  A commonly used
technique of `processing' the raw data, known as {\it pulse height 
(PH) thresholding},
takes advantage of the fact that between the two main components of
DS background, photon and particle, the former has a much narrower range
of PH.  By selecting events within this range, it is possible in principle
to suppress the particle background and improve the data signal-to-noise.
In reality, however, the method only works for point sources.  For
diffuse emission, which illuminates a large area of the detector, the
detector position dependence of this photon PH window complicates matters.
Specifically, the entire
window shifts towards higher PH as one moves from the
central region of the detector where cluster centroids are usually
located, to the outer regions where fainter radiation is detected from
large cluster radii.  

The common practice is to use a
fixed PH window everywhere, even though this window
corresponds only to the PH distribution of
photon events at or near the detector center\footnote{
A trivial way of PH thresholding the DS data, sometines employed to remove 
detector artifacts, is to apply only a lower PH threshold.  
It introduces minimal bias
to the data because off-axis photons, which are peaked at higher PH, are
no longer excluded.}.  The result is
a loss of photons, and consequently a decrease in the thresholded 
background, with increasing radius.  This is the reason why Bowyer, 
Bergh\"offer, and Korpela 1999 (hereafter 
abbreviated as BBK) recently made the erroneous conclusion that cluster EUV,
{\it including even the
radiation of the hot gas}, are mere background effects.
In fact, a map of the spatial distribution of the peak PH value of the photon
PH window (Kaastra 1999) explains precisely why BBK find much
steeper background gradients (relative to the raw data) across
the DS detector.  One could, of course, undertake the elaborate task of
applying position dependent PH windows,
as in the case of
Kaastra (1999) who also subtracted a template background obtained
from blank field pointings at random times and directions.
Even after such efforts systematic uncertainties at a known level remain,
because of the heavy
data manipulations involved and the problems related to usage of a
background template which is not obtained
{\it in situ} (see below).

The decisive way of addressing these issues is to measure the true background 
distribution of the detector region occupied by the cluster.  
We therefore scheduled an EUVE
re-observation of A2199 in February 1999, which consisted of a 47 ksec exposure
to the cluster (with the cluster center placed at 11.5 arcmin off-axis
along the detector +x direction)
immediately followed by a 11 ksec exposure of the blank sky
region (on-axis J2000 coordinates of RA = 245.523$^o$,
DEC = 40.326$^o$,
$\sim$ 2$^o$ offset from the cluster) whilst maintaining the same
roll (azimuthal) angle of the Deep Survey (DS) detector.
Within the context of the CSE this is the first direct
approach to the EUVE deep survey (DS) detector background problem, as it
involved a spatially and temporally contiguous observation - the data thus
provided represents the most relevant background map for the purpose of
subtraction.

The merits of using an offset pointing to estimate the cluster background
are clear: the offset pointing background only differed from that of the cluster
by 4\% and this should be contrasted with the $\sim$ 300 \% dynamic range of DS
raw background values for the entire EUVE mission.  Moreover, the two datasets
were found to maintain their spatial distributions of event counts and pulse
height (PH) over areas outside the cluster location, when again a broad variety
of behavior exists within the archival database.  To elaborate the latter
point, we computed the difference between the radial profiles
of the cluster and offset fields with the center of the annular system
located at 15 arcmin from boresight along the detector -x direction.
With this choice of center one avoids the effects of the
cluster emission on the profile of the cluster field, since
A2199 is $\sim$ 27 arcmin away on the other side of the detector.  Thus
one expects the subtraction to yield zero signal everywhere.
This is indeed the case, as can be seen in Figure 3a, where the data
follow a flat and vanishing profile with
$\chi^2_{red} (13)=$ 1.18, and the r.m.s. deviation is consistent
with Poisson statistics - the residual 
point-to-point systematics are $\sim$ 0.13 \% of
the pre-subtracted flux.  
In contrast, when the procedure is
repeated using the present cluster
field but another offset background field acquired 
in the same month by pointing to
a direction $\sim$ 2$^o$
away from the Virgo cluster, the subtraction was
not satisfactory, with $\chi^2_{red} (13)=$ 4.41, and a large difference
between the r.m.s. and Poisson errors 
indicating that even the
residual point-to-point systematics are at
a level higher than the random uncertainties.  Thus, e.g., a
region of apparent extended emission spanning $\sim$ 6 arcmin is
evident in the innermost 10 arcmin of the subtracted profile,
see Figure 3b.  These comparisons clearly demonstrate the
difficulties in building a background template from data obtained
contemporaneously with the cluster observation: the only
reliable template is that of a time contiguous, {\it in situ},
background.

In Figure 4a we
show the radial profile of the offset background field, now with
the annuli center placed at the detector position 
x = +11.5 arcmin occupied by the
centroid of A2199 during the cluster pointing.  The data are
consistent with a flat distribution out to $\sim$ 28 arcmin.  
Quantitatively, the difference between the
0 - 15 and 15 - 28 arcmin background levels is
1.0 $\pm$ 1.4 \%.

As to the cluster field, we show in Figure 4b the sky radial
profile of the raw data, taken from the same region of the DS detector as that
of the background (offset) pointing.  Comparison of Figure 4b with Figure 4a
suggests that cluster emission extends to $\sim$ 21 arcmin,
and one could proceed to obtain the cluster signals by
subtracting a flat background as determined from the 21 - 28 arcmin region of
the cluster field.  In the
present work, however, we undertake the most conservative approach by
performing point-to-point
subtraction of the offset background,
with propagation of background errors.  The
cluster emission profile thus obtained is shown in Figure 5a, and the resulting
soft excess profile in Figure 5b.

The rising importance of the CSE with radius, as reported in LBM (also Mittaz,
Lieu, and Lockman 1998, on A1795), is confirmed by our re-observation of A2199.
However, when compared with the original data (Figure 5b here versus Figure 1b
of LBM) the CSE beyond 10 arcmin exhibits a steeper upward trend than
previously.  The reason has to do with the new data revealing more emission in
this region.  The difference can be attributed to several features of the
re-observation: (a) the availability of offset pointing enabled us to perform a
more accurate background subtraction, (b) the exposure of the cluster to
+11.5 arcmin off-axis location where the 
DS detector has relatively higher quantum
efficiency because it has not been as exposed to bright sources
as the normal locations near boresight (Kaastra 1999, see also Figure 1), and
(c) the orientation of the DS detector, with the x-axis almost parallel to the
celestial N-S direction, is at approximate right angles to that of the first
observation - in this way the emission within 10 - 15 arcmin south-east of the
cluster center, evident in the current data, was previously missed because it
corresponded to large detector y-values, i.e. areas of lower sensitivity.  To
illustrate, we show in Figure 6 a combined image of the two observations (now
totalling an exposure of 100 ksec), adaptively smoothed in the manner described
in LBM.  The emission in the extreme south east was not evident from the EUV
contours of LBM.

The rising
trend of the CSE suggests that its origin is
a warm intracluster medium, with cooler gas residing at
larger cluster radii, and large mass implications.  Do the EUV 
signals represent the hitherto undetected baryons necessary to bridge the
gap between theoretical and observational cosmology ?  As an
example we calculated the mass of the emitting gas between cluster
radii of 12 and 15 arcmin, within the context of an equilibrium
thermal model.  The large EUV excess there, which is not accompanied
by the detection of any
soft X-ray excess in the 1/4-keV band of the ROSAT PSPC, constrains
the gas temperature to kT $\leq$ 10 eV; the gas mass is
then $\sim$ 10$^{14}$ M$_{\odot}$, on par
with that of the dark matter in the same region of the cluster 
(Siddiqui, Stewart, \& Johnstone 1998).

R. Lieu
gratefully acknowledges support from NASA's ADP and EUVE-GI programs.

\vspace{2mm}

\noindent
{\bf References}

\noindent
Bowyer, S., Bergh\"offer, T., \& Korpela, E., 1999, {\it Proc. of
workshop on `Thermal and\\ 
\indent Relativistic Plasmas in Clusters of Galaxies'},
Schl\"oss Ringberg, Germany, ed.\\
\indent H. B\"ohringer (also {\it astro-ph
9907127}). \\
\noindent
~Cen, R., \& Ostriker, J.P. 1999, {\it ApJ}, {\bf 514}, L1. \\
\noindent
~Kaastra, J.S., Lieu, R., \& Mittaz, J,P.D., 1999, {\it A \& A}, in
preparation. \\
\noindent
~Lieu, R., Mittaz, J.P.D., Bowyer, S., Lockman, F.J.,
Hwang, C. -Y., Schmitt, \\
\indent  J.H.M.M. 1996a, \it Astrophys. J.\rm, {\bf 458}, L5--7. \\
~Lieu, R., Mittaz, J.P.D., Bowyer, S., Breen, J.O.,
Lockman, F.J., \\
\indent Murphy, E.M. \& Hwang, C. -Y. 1996b, {\it Science}, {\bf 274},
1335--1338. \\
\noindent
~Lieu, R., Ip W.-I., Axford, W.I. and Bonamente, M. 1999, {\it ApJL} , \\
\indent {\bf 510}, 25--28.\\
\noindent
~Lieu, R., Bonamente, M., \& Mittaz, J.P.D. 1999, {\it ApJ}, {\bf 517}, L91. \\
\noindent
~Mittaz, J.P.D., Lieu, R., Lockman, F.J. 1998, {\it Astrophys. J.}, {\bf
498},
L17--20. \\
\noindent
~Morrison, R. and McCammon D., 1983, {\it ApJ}, {\bf 270}, 119--122.\\
\noindent
~Sarazin, C.L. 1999, {\it ApJ}, 520, 529.\\
\noindent
Siddiqui, H., Stewart, G.C., \& Johnstone, R.M., 1998,
A \& A, 334, 71. \\
\noindent
~Slezak, E., Durret, F., \& Gerbal, D. 1994, {\it AJ},
{\bf 108}, 1996.\\

\vspace{2mm}

\noindent {\bf Figure Captions}

\noindent Figure 1.  Greyscale map of the surface brightness of the DS as
obtained by co-adding the detector image of three separate pointings (details
see text) totalling an exposure of $\sim$ 85 ksec.  
This is essentially a sensitivity map of the detector
(to background events).  The contours are scaled
linearly between the maximum and minimum brightness values, which differ by 13
\%.   The normal convention adopted for the detector axes is
also indicated, with tickmarks in units of detector pixels
(13 pixels $\sim$ 1 arcmin) and with the on-axis (boresight)
position at pixel coordinates (1024,1024).

\noindent Figure 2.  The total
intensity distribution of the DS along the detector y-axis.
The dashed line represents a strip where bright point sources
affected the background - this region is
therefore not shown.  The dotted lines mark those y-limits beyond which the
profile steepens considerably.  In computing radial profiles only the y pixels
within these limits were used.  The data were obtained by merging many detector
images (see text).

\noindent Figure 3.  Radial profile of the A2199 cluster field, after
subtracting the same profile of a background field.  Both profiles
are centered at an identical highly off-axis detector position far
away from the detector region employed to observe A2199 (see text).
3a:  the background field is that of the A2199 offset pointing;
to eliminate the slight ($\sim$ 4 \%) difference between the
background of the cluster and offset pointing,
the subtraction was performed after the offset
profile was re-normalized such that the average 
asymptotic (21 - 28
arcmin) background agrees between the two fields - a procedure
adopted to extract the actual cluster signals (Figure 5).
3b: the background field is that of a very different sky direction
(see text), with the same re-normalization having been applied.
Both plots are binned in 3 arcmins to reveal the presence or
otherwise of large scale features produced by the subtraction. 
The data used are raw (i.e. no PH thresholding) and areas of
known detector artifacts were excluded from analysis.  This
statement also applies to Figures 4 and 5.
    
\noindent Figure 4.  Radial profiles of the offset (4a) and
cluster (4b) pointings of the A2199 observation.  The radius
is measured from the detector position occupied by the emission
centroid of A2199 in the cluster field.  Dotted line represents
the average brightness of the 21 - 28 arcmin region.
For perusal of the cluster emission the data here
are plotted in higher resolution than that of Figure 3, with the
offset profile displayed in 2 arcmin bins to reduce the larger random
uncertainties which result from the smaller exposure.

\noindent Figure 5.  Radial profile of the A2199 cluster EUV signal 
(5a) and soft
excess (5b), obtained by point-to-point subtraction of the background (offset)
profile, with propagation of errors from the background data and after scaling
away the $\sim$ 4 \% difference between the 21 - 28 arcmin average background
levels of the source and offset pointings.  In 5b the C-band fluxes correspond
to PSPC channels 18 - 41, and the expected Lex/B to C-band ratio of the hot ICM
emission as given by the dashed line was computed in the manner of LBM
(Figure 1b, except that the corresponding line there was lower because the
interstellar absorption code adopted here (Morrison \& McCammon 1983) 
involves removal of
less extragalactic EUV signals than before).  The lower fluxes in the inner
radii may be due to intrinsic absorption.

\noindent Figure 6.  Adaptively smoothed image of the two EUVE observations of
A2199.  Contours are cluster EUV emission
brightness in units of photons arcmin$^{-2}$ s$^{-1}$.

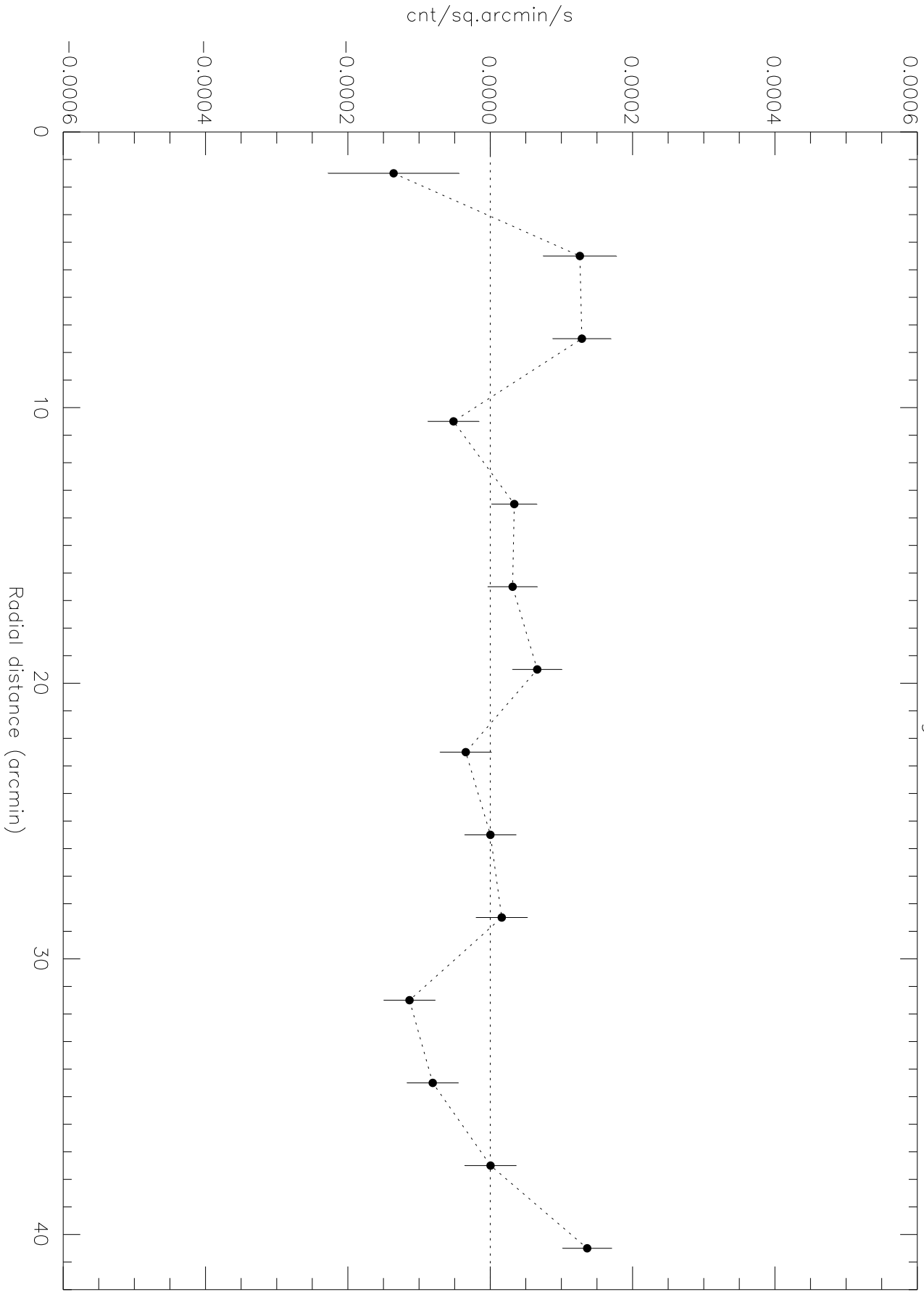
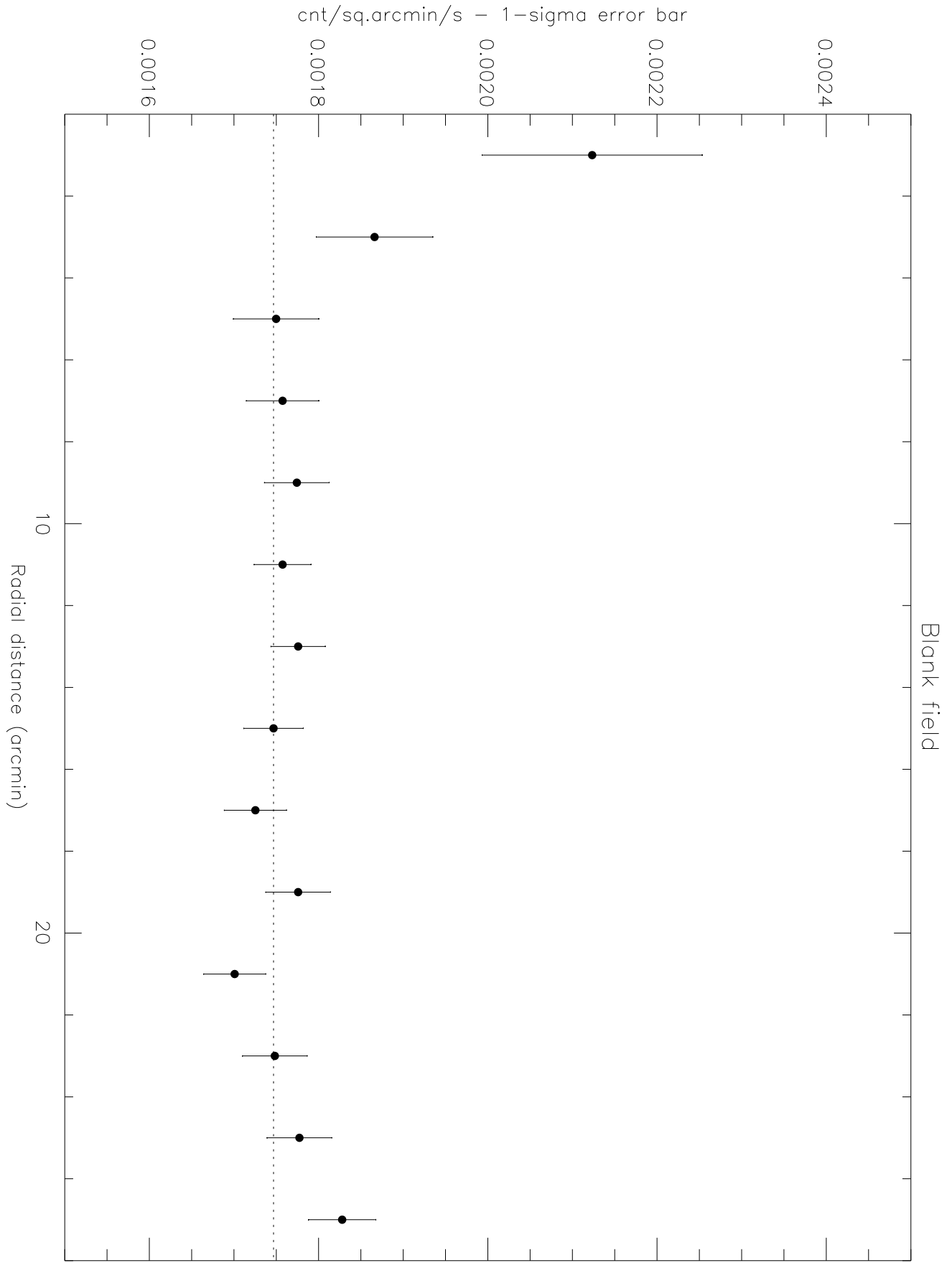
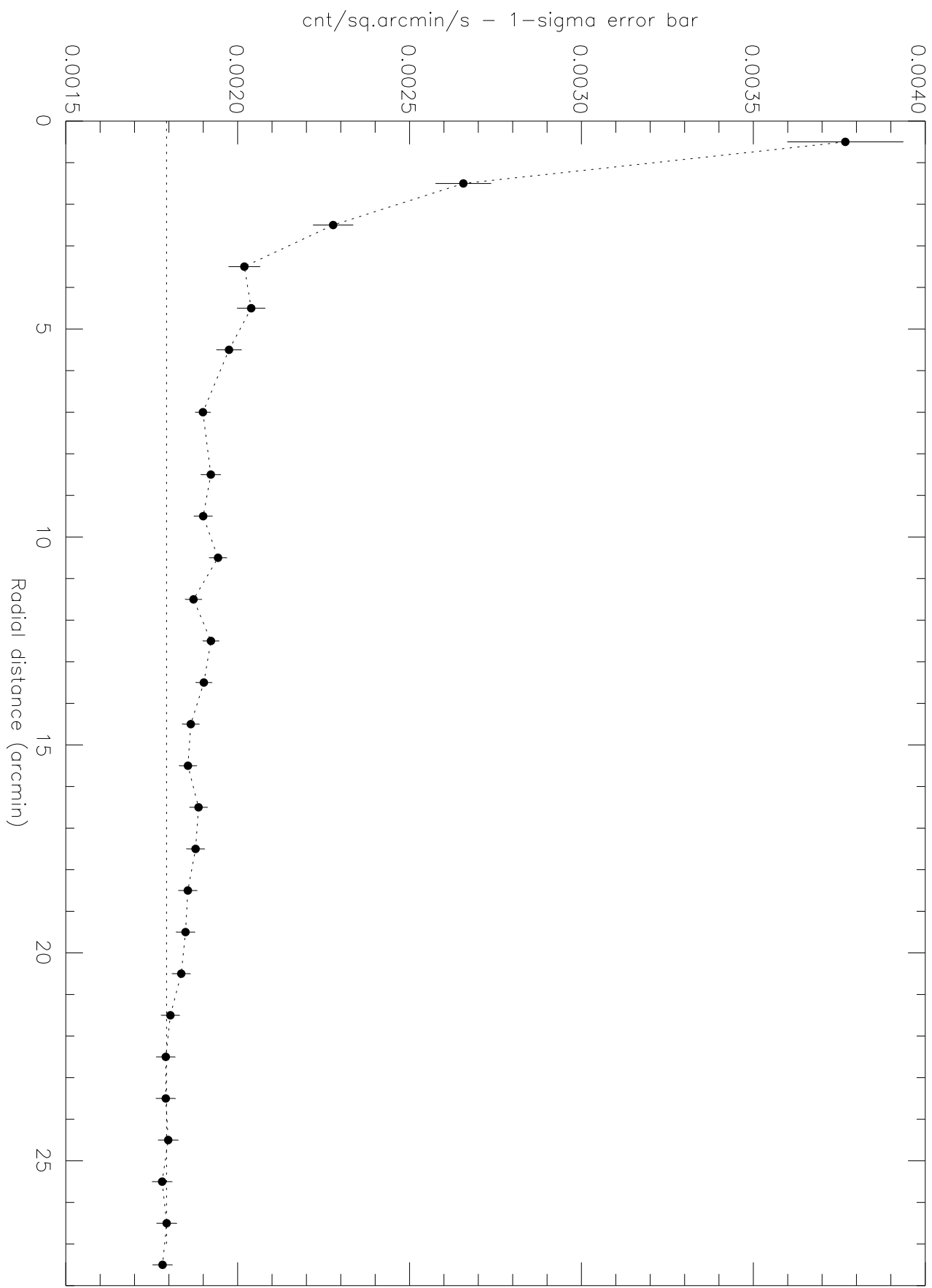
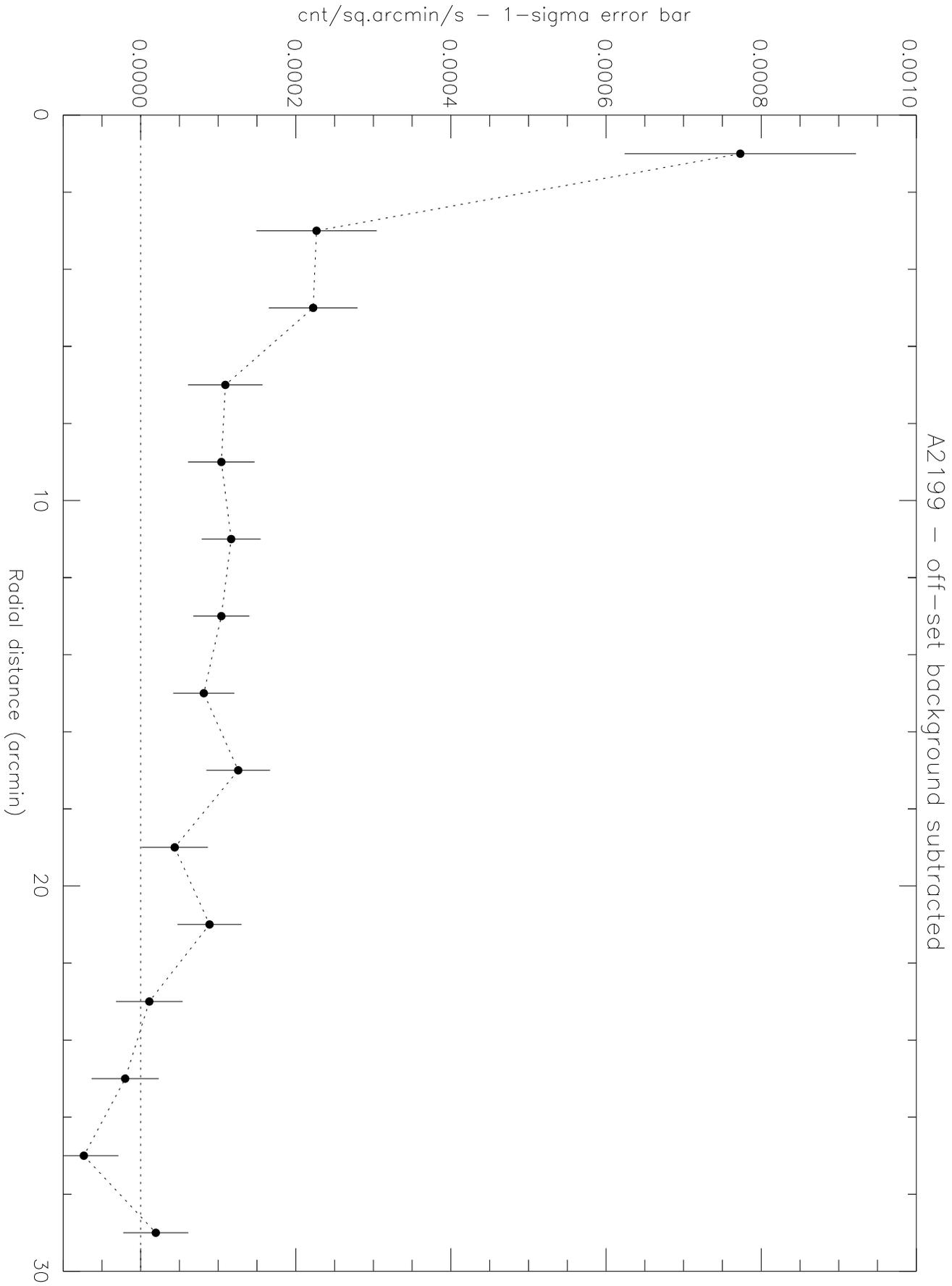
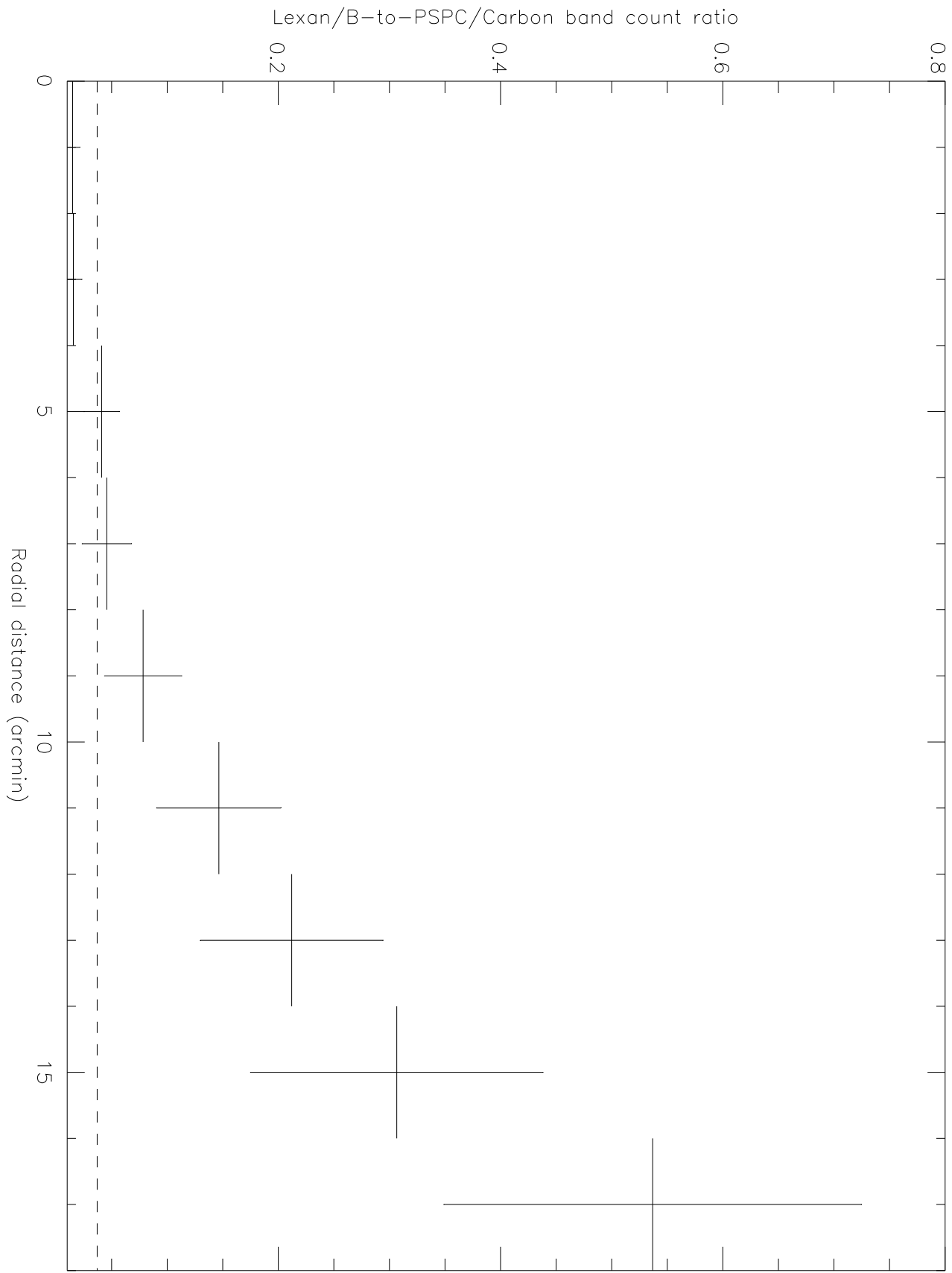
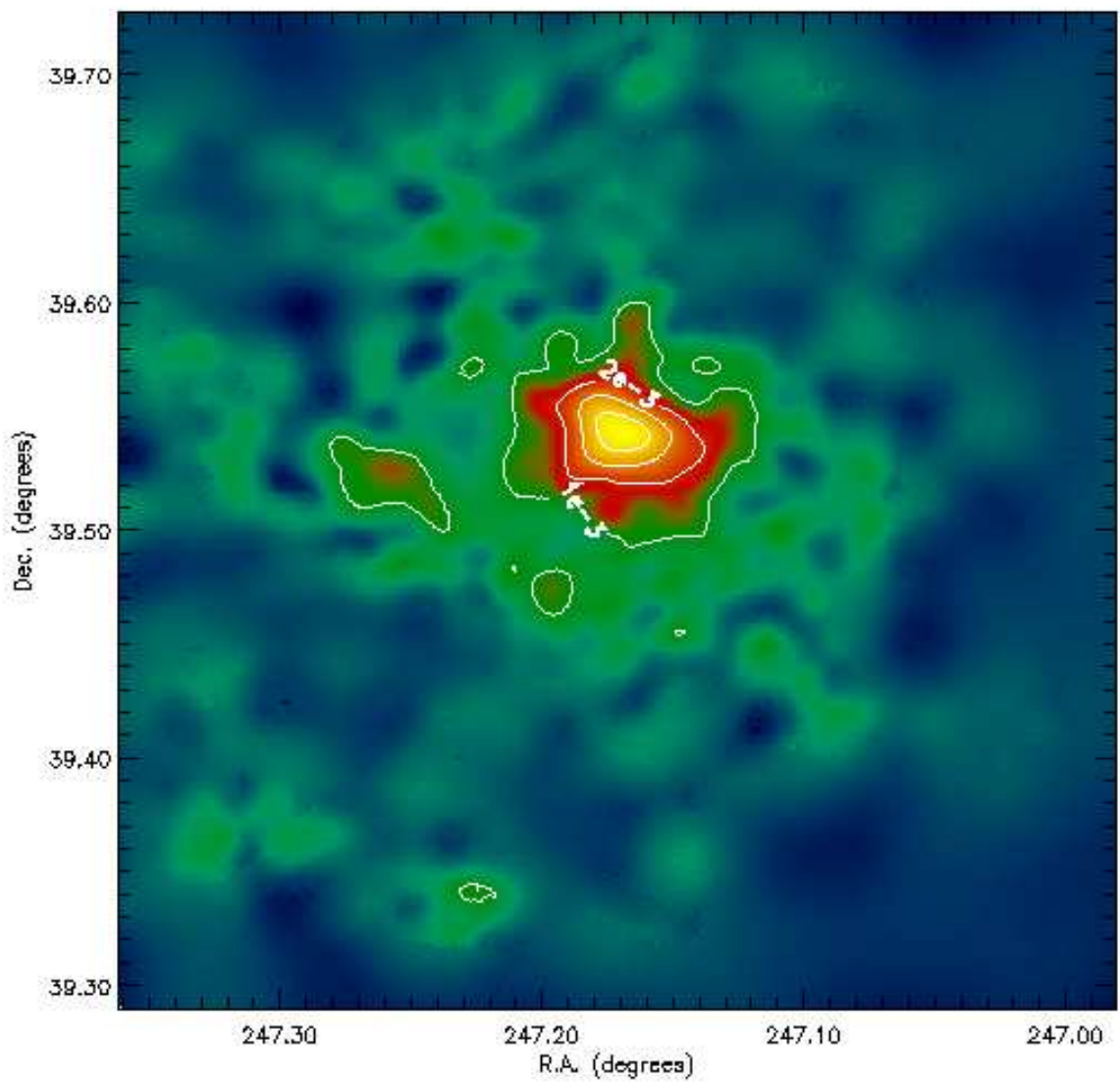
\end{document}